\documentclass[prl,twocolumn]{revtex4}
\usepackage{graphicx}
\begin{document}

\title {\Large \bf Experimental Synchronization of Spatiotemporal Chaos \\ in Nonlinear Optics}

\author{P.L. Ramazza, U. Bortolozzo and S. Boccaletti}
\affiliation{Istituto Sistemi Complessi CNR, Largo E.
Fermi, 6 I50125 Florence, Italy}

\begin{abstract}

We demonstrate that a unidirectional coupling between a pattern
forming system and its replica induces complete synchronization of
the slave to the master system onto a spatiotemporal chaotic
state.

\end{abstract}

\maketitle


In recent years, synchronization of complex systems have attracted
a great interest in the scientific community \cite{synchro}, as
well as in the literature oriented to lay audiences \cite{stro03}.
This indicates the behavior of two (or many) systems (either
equivalent or non equivalent) that adjust a common feature of
their complex dynamics due to a coupling or to a forcing.

For time chaotic systems, four types of synchronization have
mostly been studied, namely complete synchronization (CS), phase
(PS) and lag (LS) synchronization, and generalized synchronization
(GS). CS refers to a process whereby two interacting systems
perfectly link their chaotic trajectories, thus remaining in step
with each other in the course of the time \cite{complete}. GS
implies the hooking of the output of one system to a given
function of the output of the other system \cite{rul}. PS is
characterized by a locking of the phases of the two signals, also
in the absence of a substantial correlation between the two
chaotic amplitudes \cite{phase}. Finally,  LS consists in the
hooking of one system to the lagged output of the other
\cite{intermittentlag}. All these effects have been explored in
natural phenomena \cite{nature}, and laboratory experiments
\cite{exp}, and unified approaches to describe \cite{def} and
measure synchronization states have been proposed.

When the interest shifted to space-extended systems,
synchronization phenomena were shown in large populations of
coupled chaotic units and neural networks \cite{pop}, globally or
locally coupled map lattices \cite{map}, and pattern forming
systems governed by partial differential equations \cite{stc}.
Here, however, all theoretical and numerical progresses were
accompanied by a substantial lack of experimental verifications.
Precisely, CS of spatiotemporal patterns was first observed in
chemistry \cite{chemistry} for two mutually coupled
Belouzov-Zhabotinski cells, where, however, the resulting
synchronized state corresponded to the suppression of
spatiotemporal complexity and the emergence of a common spiral
behavior. Later on, LS was observed in a pair of unidirectionally
coupled nonlinear optical systems \cite{neubecker}. The evidence
here was given in terms of an  improvement in the lagged
correlation between the master and slave patterns. Finally, GS was 
demonstrated in an open loop liquid crystal light
modulator with optoelectronic feedback \cite{roy} by the use of
the so called auxiliary system method \cite{rul}.

In this Letter we report the first direct experimental evidence of
complete synchronization on unidirectionally coupled pattern forming systems. At
variance with what observed in Ref. \cite{chemistry}, the
resulting synchronized state here corresponds to a spatiotemporal
chaotic dynamics where the slave system is identically attained to
the behavior of the master system.

\begin{figure}[]
\includegraphics[width=7.5cm]{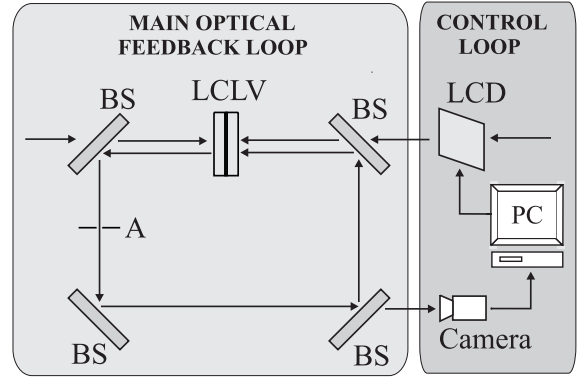}
\caption{ a) Experimental setup. Main loop: an expanded laser beam
is closed through a non-linear Kerr-like medium (liquid crystal
optical valve). BS: beam splitters; A: aperture in a Fourier plane.
Control arm: PC: personal computer; LCD: liquid
crystal display.}\label{fig:setup}
\end{figure}

The experimental setup is sketched in Fig. 1. A main optical
feedback loop (MOFL) consists of a Liquid Crystal Light Valve (LCLV) 
with optical feedback \cite{AkhmanovNoiChaos}.
The LCLV induces on the reading light a phase delay proportional to the writing
intensity (Kerr-like effect). This proportionality relation holds for all
experimental parameters used in the present investigation.
The pattern forming mechanism acts on an homogeneous pump wave of 
intensity $I_p$ sent onto the front face of the LCLV. The wave is totally
reflected and acquires a spatial phase modulation. Diffractive propagation 
along the MOFL provides conversion of phase into amplitude modulations. 
Due to the Kerr-like behaviour of the LCLV, these amplitude modulations 
are converted back to phase modulations, so that eventually a positive 
feedback establishes for some spatial frequencies, which are destabilized 
resulting in pattern formation. 

An additional electro-optic control loop
is constituted by a video-camera and a personal computer driving a
liquid crystal display (LCD). The control signal is a laser beam
that traverses the LCD before being injected into the MOFL. 
The LCD display is operating in transmission, and
encodes linearly the gray level images output by the PC, onto the
laser beam traversing it.

When the control loop is open, the dynamics of the optical beam phase $\phi
(x,y,t)$ at the LCLV output can be described by
\cite{Firth,NeubeckerOppo} $\frac{\partial \phi}{\partial t} =
-\frac{1}{\tau }(\phi -\phi_0) + D \nabla^2 \phi + \alpha I_{fb}$,
where $\phi_0$ is the working reference phase, $\tau$  the LCLV
relaxation time, $D$ a diffusion coefficient, $\alpha$ the LCLV
nonlinearity strength, and $I_{fb}(x,y,t)$ the feedback intensity
at the input plane of the FB. $I_{fb}$ is a nonlinear (and
nonlocal) function of the phase $\phi$ \cite{Firth,NeubeckerOppo}.

With increasing the pump intensity $I_p$ above the threshold
$I_{\text{thr}}$ of pattern formation, the homogeneous solution
destabilizes, and an hexagonal pattern arises. This allows
to introduce a reduced pump parameter $I \equiv
\frac{I_p}{I_{\text{thr}}}$. A further increase in $I$ above unity
leads eventually to a destabilization of hexagons in favor
of a regime of space-time chaos (STC) \cite{Firth,NeubSTC}. An
aperture $A$ (located in a Fourier plane of the MOFL) has the role
of limiting the spatial frequency bandwidth $B$ of the system,  which is 
another control parameter. Troughout the experiment here reported, 
$B$ is kept fixed at 1.5 times the diffractive spatial frequency of 
the system. This is the frequency of the hexagonal pattern 
which bifurcates at the threshold for pattern formation.  

When the control loop is closed, a fraction of $I_{fb}$ is extracted
and detected by a video-camera, which is interfaced to a personal
computer (PC) via a frame grabber. The PC processes the input
image, and sends a driving signal to the LCD, upon which a plane 
beam of intensity $I_0$ incides. The transfer
function $T(x,y,t)$ of the LCD is the sum of a constant mean transfer
coefficient $T_0$ and a modulation signal $s(x,y,t)$, which we set to be
proportional to the error signal between the actual pattern
intensity $I_{fb}$, and a desired time dependent target pattern
 $I_T(x,y,t)$ [$s(x,y,t) = -(\gamma /I_0)
\left(I_{fb}(x,y,t)-I_T(x,y,t)\right)$]. Further real time
processing performed by the PC includes the evaluation of $s$, and
the calculation of the cross-correlation between $I_{fb}$ and
$I_T$. The resulting actualization time for $s$ is of the order of 
200 to 300 ms,
to be compared with the characteristic time of the pattern
dynamics (of the order of 1 sec. for the parameters used in our
experiment). The diffractive scale of the system 
is $\sqrt {2 \lambda L} \simeq 300 \ \mu$m ($\lambda=514$ nm being
the laser wavelength, and $L=90$ mm the free propagation length in
the MOFL). On the other side, the control area is $\simeq 2000 \times
2000 \ \mu$m$^2$, and the control signal is made of $128 \times
128$ pixels. This grants us a spatial resolution of $\simeq 20$
pixels per typical pattern wavelength.

\begin{figure}[]
\includegraphics[width=9cm]{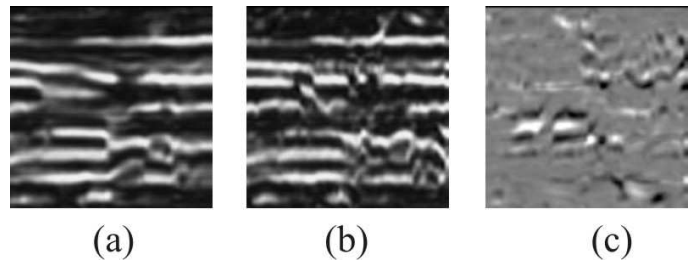}
\caption{ a) Snapshot of the master dynamics recorded with open
control loop at $I = 3$. 
a) Space (vertical)-time (horizontal)
dynamical evolution of a line of (a) the output free running signal 
from the MOFL (master dynamics); (b) the slave dynamics in 
conditions of closed control loop; (c) the synchronization error, 
corresponding to the difference of (a) and (b). The gray scale in 
(c) is such that black corresponds to negative values, white to 
positive values and gray to values close to 0.}\label{fig:synch}
\end{figure}

With the help of such real-space real-time control technique we
have recently given evidence that two dimensional stationary
target patterns with arbitrary symmetries and shapes can be
effectively and robustly stabilized within STC \cite{expnoi}.
Here, instead, we aim to demonstrating complete synchronization 
in a unidirectionally
coupled scheme between two identical systems in a regime of STC.
For this purpose, we initially let the control loop open and record
(over a time interval $T$) the free evolution of the system for
a value of $I$ at which the uncontrolled dynamics displays STC. 
A qualitative characterization of the resulting dynamics has been 
given in \cite{expnoi}. The signal consists basically of a set of closely 
packed diffractive spots, evolving in space and time in an unpredictable 
way.

After the registration of this dynamics, which we refer to as the Master
Dynamics (MD) henceforth, further free evolution is granted to the uncontrolled
system, so that after a few seconds the configuration of the MOFL output is
totally uncorrelated with the initial frame of the MD. At this point, we close
the control loop and replay the registered MD as the target pattern
$I_T(x,y,t)$. In this way, we are implementing a unidirectional coupling scheme
between two {\it identical} systems starting from fully uncorrelated initial
conditions. By repeating the replaying procedure of the MD with increasing
values of $\gamma$ (hereinafter called the coupling strength parameter), we
eventually observe full synchronization between the controlled output of the
MOFL [the slave dynamics (SD)] and the MD.

Complete synchronization between SD and MD is shown in Fig. 2,
reporting the space (vertical)-time (horizontal) dynamical
evolution of the central vertical line of pixels for the 
master dynamics (2a), slave dynamics (2b) and the difference between 
the two, (2c). Fig, 2 is taken at 
$\gamma=0.8$, and shows how the SD closely follows the MD at any
time during control. As visible in fig, 2c, the synchronization error 
$E(x,y,t)= I_{fb}(x,y,t)-I_T(x,y,t)$
is close to vanishing nearly always and 
everywhere in time as a result of the complete synchronization
process. 

Notice that the final CS state is here realized within a
full STC regime, at variance with what reported in Ref.
\cite{chemistry} for a bidirectional coupling between two
excitable media, where the emergence of synchronization was
associated with the suppression of space-time chaos in the system.

\begin{figure}[]
\includegraphics[width=9cm]{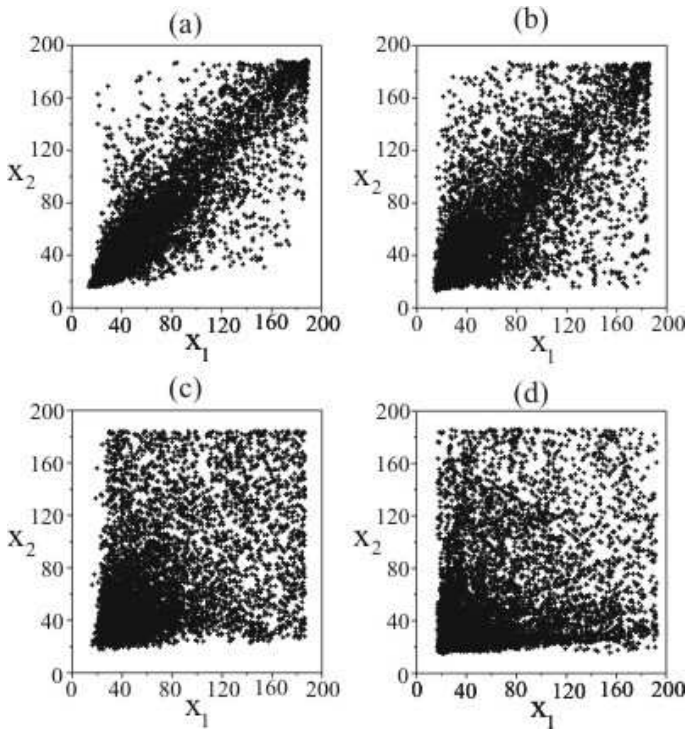}
\caption{Experimental point distributions in the
$(x_{\text{SD}},x_{\text{MD}})$ plane (see text for definition of
the variables). The synchronization manifold is represented by the
diagonal line $x_{\text{SD}}=x_{\text{MD}}$. (a) $I=2, \ \gamma=0.8$,
b) $I=3, \ \gamma=0.8$, c) $I=2, \ \gamma=0$, d) $I=3, \
\gamma=0$.}\label{fig:manifold}
\end{figure}

An independent way of visualizing the emergence of CS is to pick
randomly a set of $N$ points [$(x_i,y_i); \ i=1, \ldots , N$] in
both the MD and the SD, and to plot the variable $x_{\text{SD}}
(t)= \left\{ I_{fb}(x_i,y_i,t), \ \  i=1, \ldots , N \right\}$
{\it vs.} the corresponding variable $x_{\text{MD}} (t)= \left\{
I_T(x_i,y_i,t), \ \ i=1, \ldots , N \right\}$. The more the
distribution of points in the $(x_{\text{SD}},x_{\text{MD}})$
plane approaches the synchronization manifold (the diagonal line
$x_{\text{SD}}=x_{\text{MD}}$), the cleaner a CS state is set in
our system at all times. Experimental results are reported in Fig.
3 for two values of the coupling strength $\gamma$. Precisely,
Fig. 3a (3c) shows the distribution of points in the
$(x_{\text{SD}},x_{\text{MD}})$ plane for $I=2$ and $\gamma=0.8$
($\gamma=0$), while Fig. 3b (3d) refers to the same situation for
$I=3$ and $\gamma=0.8$ ($\gamma=0$). In both cases, it is apparent
that increasing the coupling strength induces a point distribution
much closer to the diagonal line than the uncoupled case, giving
evidence that a CS state has arisen in the experiment.

\begin{figure}[]
\includegraphics[width=6cm]{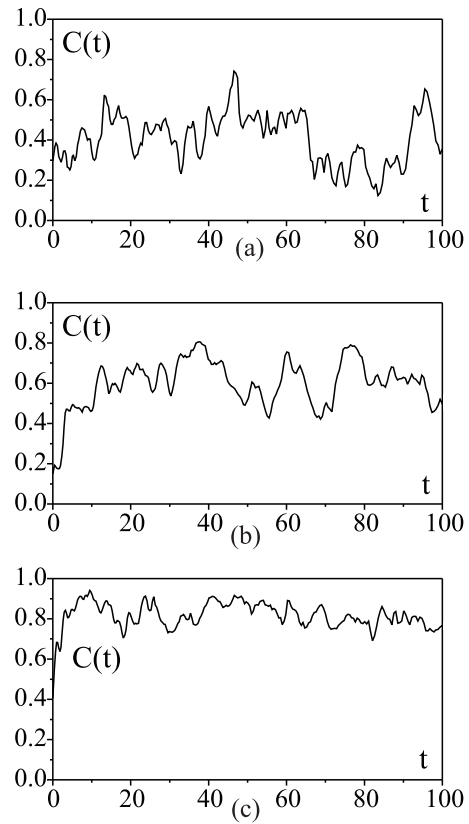}
\caption{Cross correlation function $C(t)$ (see text for
definition) {\it vs.} time during some synchronization trials for
$I=3$. a) $\gamma =0.24$, b) $\gamma =0.40$, c) $\gamma =0.8$. In
all cases $t=0$ indicates the instant at which the MD starts to be
replayed in the control loop.}\label{fig:corrvstime}
\end{figure}

A more quantitative measurement of CS can be given by monitoring
the behavior of the time dependent cross-correlation function
$C(t)=<I_{fb}(\textbf{r},t)\cdot I_T(\textbf{r},t)>_{\textbf{r}}$
between the instantaneous patterns in the SD and MD during the
synchronization process ($< \ldots >_{\textbf{r}}$ denotes here a
spatial average in the plane ${\textbf{r}} \equiv (x,y)$). $C(t)$
is by definition vanishing for linearly uncorrelated systems,
whereas $C(t) \sim 1$ for fully synchronized dynamics.

Fig. 4 reports the temporal behavior of $C(t)$ for $I=3$ and for
three different values of the coupling strength [a) $\gamma =0.24$,
b) $\gamma =0.4$, and c) $\gamma =0.8$]. In all horizontal axes,
$t=0$ has been taken as the instant at which the MD starts to be
replayed in the control loop. A first important observation is the
cross correlation starts from a non vanishing value at $t=0$. This
is because the uncontrolled STC dynamics has a non zero mean
field, as it can be appreciated from inspection of Fig. 2a-2b.
Namely, the uncoupled MD and SD have a certain degree of "phase
rigidity", i.e., even if there are chaotic fluctuations, bright
(dark) areas remain more or less bright (dark) for most of time.
Similar properties have been observed experimentally and discussed
in various other cases of space extended systems giving rise to
STC dynamics \cite{vari}.

At low values of the coupling strength ($\gamma=0.24$), no
synchronization is set in the system. This is visible in Fig. 4a,
where $C(t)$ experiences large fluctuations in time around a mean
value not substantially different from the initial correlation
level. For intermediate coupling strengths ($\gamma=0.4$ in Fig.
4b), a partial synchronization emerges in the system after a
transient time, though several deviations of the SD from the MD
still remain, reflected by the rather large fluctuations around
the asymptotic value of $C(t)$ visible in Fig. 4b. Finally, a
further increase in $\gamma$ induce a full CS in the system (see
the case $\gamma=0.8$ in Fig. 4c), where the asymptotic value of the
cross correlation approaches unity, and the residual fluctuations
in $C(t)$ shrink considerably. Notice that, as $\gamma$ increases,
the transient time before reaching CS decreases. The global
picture depicted in Fig. 4 confirms that CS here is a threshold
phenomenon, as it was introduced originally for time-chaotic
systems \cite{complete}.

\begin{figure}[h!]
\includegraphics[width=7.5cm]{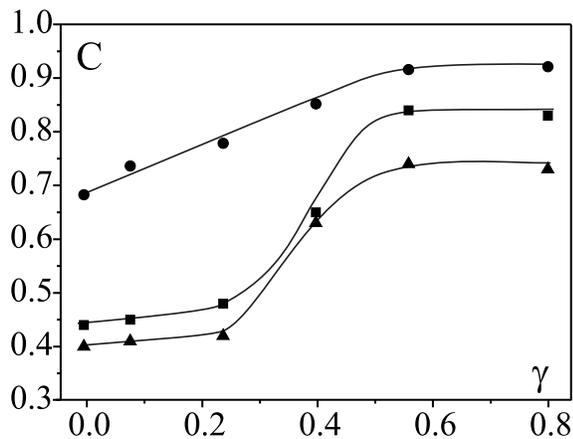}
\caption{Time average cross correlation function $C$ (see text for
definition) {\it vs.} the coupling strength $\gamma$ for $I=2$
(circles), $I=3$ (squares), and $I=4$ (triangles). In all cases,
the solid lines connecting points are drawn as guides for the
eyes.}\label{fig:averagecorr}
\end{figure}

The global scenario of observed CS is illustrated in Fig. 5, where
the time average of the cross correlation function $C =
<C(t)>_T$ ($<\ldots>_T$ here denotes a further time average
over the full time interval $T$) is reported {\it vs.} the
coupling strength at different values of pump intensities $I$. In
all cases, our proposed coupling scheme is effective in inducing
CS, as $C$ considerably increases with $\gamma$ with respect to
the corresponding uncoupled values. An increase in the pump
intensity $I$ leads to a progressive deterioration of CS,
reflected by smaller asymptotic values of $C(t)$. 

In conclusion, we have demonstrated that a unidirectional coupling
between two identical pattern forming systems induces complete
synchronization of the slave to the master dynamics, with a
resulting synchronized behavior corresponding to a spatiotemporal
chaotic state. Though realized with a nonlinear optical
experiment, the coupling scheme used (based on a real-space
real-time control of a recorded and replayed space-time chaotic
dynamics) can in principle be implemented in many other physical
and chemical pattern forming systems. Furthermore, the great
flexibility offered by the proposed coupling scheme can be
exploited to drive the slave system onto a generic desired
dynamics with arbitrary symmetries and shapes in space, as well as
arbitrary behaviors in time.

Work partly supported by  MIUR-FIRB project n. RBNE01CW3M-001.



\begin{thebibliography}{99}


\bibitem{synchro}
S. Boccaletti, J. Kurths, G. Osipov, D. Valladares and C. Zhou,
Phys. Rep.  {\bf 366}, 1, (2002).

\bibitem{stro03}
S. Strogatz, "Sync: The Emerging Science of Spontaneous Order",
 Hyperion Press, 2003.

\bibitem{complete}
H. Fujisaka and T. Yamada, Prog. Theor. Phys. {\bf 69}, 32 (1983);
L.M. Pecora and T.L. Carroll, Phys. Rev. Lett. {\bf 64}, 821
(1990).

\bibitem{rul}
L. Kocarev and U. Parlitz, Phys. Rev. Lett. {\bf 76}, 1816,
(1996).

\bibitem{phase}
M.G. Rosenblum, A.S. Pikovsky and J. Kurths, Phys. Rev. Lett. {\bf
76}, 1804 (1996).

\bibitem{intermittentlag}
M.G. Rosenblum, A.S. Pikovsky and J. Kurths, Phys. Rev. Lett. {\bf
78}, 4193 (1997); S. Boccaletti and D.L. Valladares, Phys. Rev.
{\bf E62}, 7497 (2000).


\bibitem{nature}
C. Schafer, M.G. Rosemblum, J. Kurths and H.H. Abel, Nature {\bf
392}, 239 (1998);  G. M. Hall, S. Bahar and D.J. Gauthier, Phys.
Rev. Lett. {\bf 82}, 2995 (1999).

\bibitem{exp}

L.M. Pecora and T.L. Carrol, Phys. Rev. Lett. {\bf 64}, 821 (1990); 
K.M. Cuomo and A. V. Oppenheim, Phys. Rev. Lett. {\bf 71}, 65 (1993); 
R. Roy and K.S. Thornburg, Phys. Rev. Lett. {\bf 72}, 2009 (1994);
D. Maza, A. Vallone, H. Mancini and S. Boccaletti, Phys. Rev. 
Lett. {\bf 85}, 5567 (2000); 
H.B. Pedersen et Al., Phys. Rev. Lett. {\bf 87}, 055001 (2001).


\bibitem{def}
R. Brown and L. Kocarev, Chaos {\bf 10}, 344 (2000); S.
Boccaletti, Louis M. Pecora, A. Pelaez, Phys. Rev. {\bf E63},
066219 (2001).

\bibitem{pop}
S. H. Strogatz, R.E. Mirollo and P.C. Matthews, Phys. Rev. Lett.
{\bf 68}, 2730 (1992); D.H. Zanette, Phys. Rev. {\bf E55}, 5315
(1997); V. N. Belykh, I.V. Belykh and M. Hasler, Phys. Rev. {\bf
E62}, 6332 (2000).

\bibitem{map}
V. N. Belykh and E. Mosekilde, Phys. Rev. {\bf E54}, 3196 (1996);
A. Pikovsky, O. Popovych and Yu. Maistrenko, Phys. Rev. Lett. {\bf
87}, 044102 (2001).

\bibitem{stc}
H. Gang and QuZhilin, Phys. Rev. Lett. {\bf 72}, 68 (1994); L. Kocarev,
Z. Tasev and U. Parlitz, Phys. Rev. Lett. {\bf 79}, 51 (1997); S.
Boccaletti, J. Bragard, F.T. Arecchi and H. Mancini, Phys. Rev.
Lett. {\bf 83}, 536 (1999).

\bibitem{chemistry}
M. Hildebrand, J. Cui, E. Mihaliuk, J. Wang and K. Showalter,
Phys. Rev. {\bf E68}, 026205 (2003).

\bibitem{neubecker}
R. Neubecker and B. G\"utlich, Phys. Rev. Lett. {\bf 92}, 154101
(2004).

\bibitem{roy}
E.A. Rogers, R. Kalra, R.D. Schroll, A. Uchida, D.P. Lathrop and
R. Roy, Phys. Rev. Lett. {\bf 93}, 084101 (2004).

\bibitem{AkhmanovNoiChaos}
S.A. Akhmanov, M.A. Vorontsov and V. Yu Ivanov, JETP Lett. {\bf
47}, 707 (1988).

\bibitem{Firth}
G. D'Alessandro  and W.J. Firth, Phys. Rev. Lett. {\bf 66} 2597
(1991).

\bibitem{NeubeckerOppo}
R. Neubecker , G.L. Oppo, B. Thuering and T. Tschudi, 1995 Phys.
Rev. {\bf A52}, 791.

\bibitem{NeubSTC} R. Neubecker, B. Thuring, M. Kreuzer and T. Tschudi,
Chaos, Solitons and Fractals {\bf 10}, 681 (1999).

\bibitem{expnoi}
L. Pastur, L. Gostiaux, U. Bortolozzo, S. Boccaletti, and P. L.
Ramazza, Phys. Rev. Lett. {\bf 93}, 063902 (2004).

\bibitem{vari}
B.J. Gluckman, P. Marcq, J. Bridger and J.P. Gollub, Phys. Rev.
Lett. {\bf 71}, 2034 (1993); Li Ning, Y. Hu, R.E. Ecke and G.
Ahlers, Phys. Rev. Lett. {\bf 71}, 2216 (1993); E. Bosch, H.
Lambermont and W. van de Water, Phys. Rev. {\bf E49}, R3580
(1994); S. Rudroff and I. Rehberg, Phys. Rev. {\bf E55}, 2742
(1997).

\end{thebibliography}
\end{document}